\documentclass[pra,twocolumn,nobibnotes,superscriptaddress]{revtex4-1}
\usepackage{graphicx}
\usepackage{multirow,bigdelim}
\usepackage{bm}
\usepackage{amsmath}
\usepackage{lipsum}
\usepackage[breaklinks=true]{hyperref}
\usepackage{float}

\usepackage[english]{babel}
\usepackage{blindtext}

\usepackage{color}

\hypersetup{colorlinks=true,
pdftitle={Voigt spectroscopy paper},
pdfauthor={James Keaveney, Francisco S. Ponciano-Ojeda},
linkcolor=blue,
citecolor=blue,
urlcolor=blue}

\begin{document}

\title{Quantitative optical spectroscopy of $^{87}$Rb vapour in the Voigt geometry in DC magnetic fields up to 0.4~T}
\date{\today}

\author{James Keaveney}
\email{james.keaveney@durham.ac.uk}
\affiliation{Joint Quantum Centre (JQC) Durham-Newcastle, Department of Physics, Durham University, South Road, Durham, DH1 3LE, United Kingdom}
\author{Francisco S. Ponciano-Ojeda}
\email{francisco.s.ponciano-ojeda@durham.ac.uk}
\affiliation{Joint Quantum Centre (JQC) Durham-Newcastle, Department of Physics, Durham University, South Road, Durham, DH1 3LE, United Kingdom}
\author{Stefan M. Rieche}
\affiliation{Joint Quantum Centre (JQC) Durham-Newcastle, Department of Physics, Durham University, South Road, Durham, DH1 3LE, United Kingdom}
\author{Mark J. Raine}
\affiliation{Superconductivity Group, Department of Physics, Durham University, South Road, Durham, DH1 3LE, United Kingdom}
\author{Damian P. Hampshire}
\affiliation{Superconductivity Group, Department of Physics, Durham University, South Road, Durham, DH1 3LE, United Kingdom}
\author{Ifan G. Hughes}
\affiliation{Joint Quantum Centre (JQC) Durham-Newcastle, Department of Physics, Durham University, South Road, Durham, DH1 3LE, United Kingdom}

\begin{abstract}
We present a detailed spectroscopic investigation of a thermal $^{87}$Rb atomic vapour in magnetic fields up to 0.4~T in the Voigt geometry. We fit experimental spectra with our theoretical model \textit{ElecSus} and find excellent quantitative agreement, with RMS errors of $\sim 0.3$\%. We extract the magnetic field strength and the angle between the polarisation of the light and the magnetic field from the atomic signal and find excellent agreement to within $\sim 1$\% with a commercial Hall probe. Finally, we present an investigation of the relative sensitivity of this technique to variations in the field strength and angle with a view to enabling atom-based high-field vector magnetometry.
\end{abstract}

\maketitle


\section{Introduction}

Since the redefinition of the second based on the Cs hyperfine ground-state energy splitting in 1967~\cite{Terrien1968}, atomic spectroscopy has been established as the fundamental building-block of high precision measurement systems.
In recent years, spectroscopic techniques have been extended to a more general family of atom-based applications.
These applications include atomic clocks~\cite{Campbell2011,Ludlow2015}, determination of fundamental constants~\cite{Truong2015}, electrometry in DC~\cite{Osterwalder1999,Thiele2015}, microwave~\cite{Sedlacek2013,Kumar2017,Kumar2017a} and THz fields~\cite{Wade2016b,Wade2018}, and near-field imaging of microwave electric circuits~\cite{Horsley2015,Horsley2016}.
There is also the development of new sensors, including magnetometers~\cite{Budker2007} with optically-pumped atomic vapours which have impact across a wide variety of disciplines. Recent examples include explosives detection~\cite{Lee2006}, gyroscopes~\cite{Donley2009}, medical imaging of the heart~\cite{Bison2009,Alem2015} and brain~\cite{Sander2012,Boto2017}, microfluidics~\cite{Xu2008} and measurements on spin-active solid-state systems~\cite{Arnold2016}.
Applications also include optical devices including an atom-based optical isolator~\cite{Weller2012b}, a dichroic beamsplitter based on the Faraday effect~\cite{Abel2009a}, and extremely narrow-band optical band-pass filters~\cite{Kiefer2014,Zentile2015g,Zentile2015a,Rotondaro2015,Portalupi2016} which have also recently been used as an intracavity frequency-selective element in a laser system~\cite{Keaveney2016c}.
Together, these tools demonstrate an in-depth understanding and use of the atomic physics involved in externally applied magnetic fields up to $\sim 1$~T.

Atom-based measurement systems for large fields above 1~T have received significantly less attention.
In large pulsed magnetic fields, the Zeeman splitting in alkali-metal atoms has been used to measure fields up to 58~T~\cite{Ciampini2017a,George2017} with non-destructive field production, and with destructive techniques up to 200~T~\cite{Gomez2014} and 500~T~\cite{Garn1966} using the sodium D-line splitting.
This research has important applications in high energy-density science for magnetically imploded inertial fusion~\cite{Gomez2014} and a variety of other fundamental physics investigations~\cite{Battesti2018a}.
In DC fields above 1~T, Nuclear Magnetic Resonance (NMR) is currently the gold-standard for high accuracy measurements and commercial devices claim measurement precision at the 10 parts-per-billion level~\footnote{\url{https://www.metrolab.com/products/pt2026/}, accessed September 28, 2018}.
There is some work \cite{Sargsyan2015c,Sargsyan2012} which would serve as a precedent for further experiments directed at making measurements in high DC fields with a view to realising atom-based sensors in this field range.

Atomic spectroscopy of thermal vapours in magnetic fields is extremely well-understood, and the absolute absorption in both low and high-density vapours~\cite{Siddons2008b,Weller2011a}, and magneto-optic effects~\cite{Siddons2009a,Siddons2009b,Siddons2010,Weller2012} of thermal vapours of alkali-metal atoms have already been extensively studied.
In fields below 10~mT, the Larmor precession frequency is sufficiently low to be measured directly~\cite{Budker2007}. 
In fields of order 1~T, the nuclear and electronic spins decouple (the hyperfine Paschen-Back (HPB) regime), and the field can be measured from the characteristically symmetric absorption spectra~\cite{Weller2012a,Sargsyan2014} and dispersion properties that makes extracting the medium's absorption coefficient the preferred approach.
To date, most work has been done in the Faraday configuration, where the $\vec{k}$ vector of the light is parallel to the external field $\vec{B}_{\rm{ext}}$ and only the magnitude of the field can be measured~\cite{Zentile2014a}.
Here we explore the much less well studied absorption spectrum in the Voigt geometry in a DC field, where the external magnetic field is perpendicular to the direction of light propagation and not only the magnitude of the field but its direction can also be measured.
This work may enable the development of new vector magnetometers for use in magnet technology for mapping high magnetic fields.
High field applications include accelerator magnet technology for advanced high-energy physics~\cite{FCC2014}, magnets for particle beam therapy~\cite{Wan2015,Tian2018}, superconducting undulators to improve the performance of light sources~\cite{Hwang2011}, very high field magnets for future fusion reactors~\cite{Aymar2002,TSLee2015} and research magnets.
Very high accuracy atom based vector magnetometry may also enable the development of extremely small sensors (that do not require leads), non-invasive magnetometry, for applications where measurements of magnetic fields without significantly perturbing them is required, and a new type of generic magnetic characterisation tool for measuring the properties of anisotropic magnetic and superconducting materials in high fields~\cite{NRC2013}.

Here we present careful analysis of spectroscopic measurements of atomic absorption spectra of $^{87}$Rb over a range of DC magnetic field strengths up to $0.4$~T in the Voigt geometry.
By fitting to our theoretical model \textit{ElecSus} that describes atomic absorption~\cite{Zentile2015b,Keaveney2017a} we extract the magnetic field strength and compare it against experimental values.
We show that we can extract, given a known input light polarisation, the magnetic field and direction.
Finally, we calculate and evaluate the relative spectral sensitivity to changes in field strength and direction.

The rest of the manuscript is organised as follows. In section~\ref{sec:theory} we summarise the theoretical model we have used in our analysis; section~\ref{sec:expt} discusses our experimental approach; in section~\ref{sec:spectra} we discuss the spectra of $^{87}$Rb in large magnetic fields and compare experimental data to our model to extract a magnetic field strength; and finally, in section~\ref{sec:sensitivity} we explore potential areas of the spectrum which exhibit especially high or low magnetic field sensitivities.

\section{Theoretical model}
\label{sec:theory}

The theoretical model we use for fitting our data, {\it ElecSus}, is described in detail in refs.~\cite{Zentile2015b,Keaveney2017a}.
Here we summarise only the main points.
The model calculates the complex electric susceptibility of the atomic medium as a function of the optical frequency detuning, which can be written for a two-level atom as
\begin{align}
\label{eqn:CompChi}
	\chi(\Delta_{j}) = \chi_{0}^{j} \frac{-1}{\Delta_{j} + i(\Gamma_{\rm{Nat}}/2)}.
\end{align}
Here $\chi_{0}^{j}$ is a constant factor for the $j$-th transition that depends on the dipole matrix element, $\Gamma_{\rm{Nat}}$ is the natural Lorentzian linewidth for the atom, $\Delta_{j}$ is the optical frequency detuning defined by $\hbar\Delta_{j} = (\hbar\omega_{\rm{laser}} - \hbar\omega_{j})$ with $\omega_{\rm{laser}}$ the angular frequency of the laser and $\omega_{j}$ the angular frequency of the atomic transition.
Using matrix methods we construct the atomic states in the $m_L, m_S, m_I$ basis, accounting for the internal energy levels (fine and hyperfine structure) and the interaction with the external magnetic field via the Zeeman effect.
From the matrices we calculate the transition energies and absolute line strengths which will give a different factor $\chi_{0}^{j}$ for each transition.
Finally, the Doppler effect is included by considering the effect of the atomic velocities $v$ on the detuning, $\Delta_{j} \to \Delta_{j} - k v$, where $k$ is the light propagation vector of the light. As such, the susceptibility can be expressed as
\begin{align}
\label{eqn:ChiDoppler}
	\chi_{v}(\Delta_{j}) = \int_{-\infty}^{\infty}f(v)\chi(\Delta_{j} - kv)dv,
\end{align}
where $f(v)$ is the Maxwell-Boltzmann velocity distribution of the atoms, $f(v) \propto \exp{(-mv^{2}/2k_{\rm{B}}T)}$.
This results in a Voigt profile, a convolution of a Lorentzian with a homogeneous linewidth increased by presence of buffer gas, $(\Gamma_{\rm{Nat}} + \Gamma_{\rm{Buf}})$, from the imaginary part of (\ref{eqn:CompChi}) and the Gaussian from the Maxwellian velocity distribution in (\ref{eqn:ChiDoppler}) with a spectral width $\Gamma_{\rm{Doppler}}$, that describes the lineshape of the absorption of the atomic transitions as a function of detuning.

For a given global detuning $\Delta$, the susceptibility $\chi_{\rm{T}}$ is the sum over all the transitions, where each transition has an effective detuning $\Delta - \Delta_{j}$:
\begin{align}
\label{eqn:Chi_Total}
	\chi_{\rm{T}}(\Delta) = \sum_{j} \chi_{v}(\Delta - \Delta_{j}).
\end{align}
The susceptibility can be calculated separately for the $\sigma^\pm$ and $\pi$ transitions by taking the appropriate dipole matrix elements in the definition of $\chi_{0}^{j}$.
The result from equation (\ref{eqn:Chi_Total}) can then be used to write the absorption coefficient $\alpha$ with which the total absorption by the atomic medium can be examined
\begin{align}
\label{eqn:S0_def}
	S_{0}(\Delta) = I(z=&L)/I_{0} =  \exp[-\alpha(\Delta) L], \\
	\alpha(\Delta) =&\,k\rm{Im}[\chi_{\rm{T}}(\Delta)]\nonumber
\end{align}
where $k$ is the propagation vector of the system and $L$ is the length of the cell containing the atomic medium.

\begin{figure}[t]
\includegraphics[width=0.5\columnwidth]{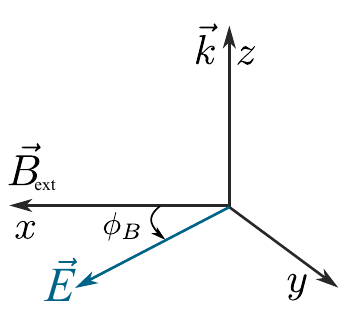}
\caption{Geometry of the externally applied field and the light used in the Voigt geometry, where $\vec{k}\perp \vec{B}_{\rm{ext}}$. The linearly polarised light has an electric field vector $\vec{E}$ that lies in the $x-y$ plane and makes an angle $\phi_{B}$ with respect to $\vec{B}_{\rm{ext}}$. This angle determines the parallel and perpendicular components of the $\vec{E}$ which in turn are associated with the different refractive indices in the atomic medium and $\pi$ and $\sigma^{\pm}$ transitions it drives.}
\label{fig:geometry}
\end{figure}

Once the susceptibility has been calculated, the wave equation is solved for the medium to find the two propagating eigenmodes.
Each eigenmode is associated with a complex refractive index, which couples to the atomic transitions in a distinct way.
The exact coupling depends on the geometry of the system.
The present experiment is set up in the Voigt geometry, in which the externally applied magnetic field vector $\vec{B}_{\rm{ext}}$ is perpendicular to the light wavevector $\vec{k}$, as shown in figure~\ref{fig:geometry}.
We constrain the external magnetic field vector to lie along the Cartesian $x$-axis, while the laser beam propagates along the $z$-axis and assume we have a linearly polarised plane-wave so its polarisation lies in the $x-y$ plane.
The angle, $\phi_{B}$, that the $E$-field of the light makes with the $x$-axis, defines the angle of polarisation and determines the relative coupling to the atomic transitions.
The two refractive indices of the atomic medium are both associated with the propagation of linearly polarised light; light polarised with its electric field parallel to the external magnetic field drives $\pi$ transitions, while light polarised with the electric field perpendicular to the external magnetic field drives both $\sigma^\pm$ transitions.
The relative phase between the two polarisation components is unimportant in the Voigt geometry - i.e. circularly polarised light of either handedness couples to the atoms in the same way as linearly polarised light with equal $x$ and $y$ components ($\phi_{B}=(2n-1)\pi/4$ with $n$ an integer).
Such considerations, including the oscillatory nature of light, mean that the direction of the applied field cannot be derived from a single measurement, meaning that equivalent solutions are found for changes in $\phi_{B}$ of $\pi$ radians.

Each eigenmode propagates through the medium with its associated refractive index.
In general these two refractive indices are different, and since the refractive indices are complex, the medium is both dichroic and birefringent.
To calculate the electric field after propagating through the medium, we transform into the eigenbasis coordinate system (in the Voigt geometry this is simply the Cartesian basis parallel and perpendicular to the external magnetic field vector), propagate each index $n_i$ for a distance $L$ by multiplying by $e^{{\rm i}n_i k L}$, and transform back to the lab coordinates which are most relevant.
We then analyse the output via Stokes polarimetry~\cite{Weller2012}, which provides a convenient set of parameters easily amenable to laboratory measurements that only require measuring the intensity in sets of orthogonal polarisation bases.
The Stokes parameter $S_0$ (see equation (\ref{eqn:S0_def})) represents the total transmitted intensity, and is therefore independent of the measurement basis.
This work has used the publicly-available version of \textit{ElecSus}, without any modifications, that has been extensively validated as seen in~\cite{Zentile2015b,Keaveney2017a}.
We find it remains valid over the extended experimental parameter space presented here for all fits and simulated datasets in this work.

%
%
\begin{figure}[t]
\includegraphics[width=\columnwidth]{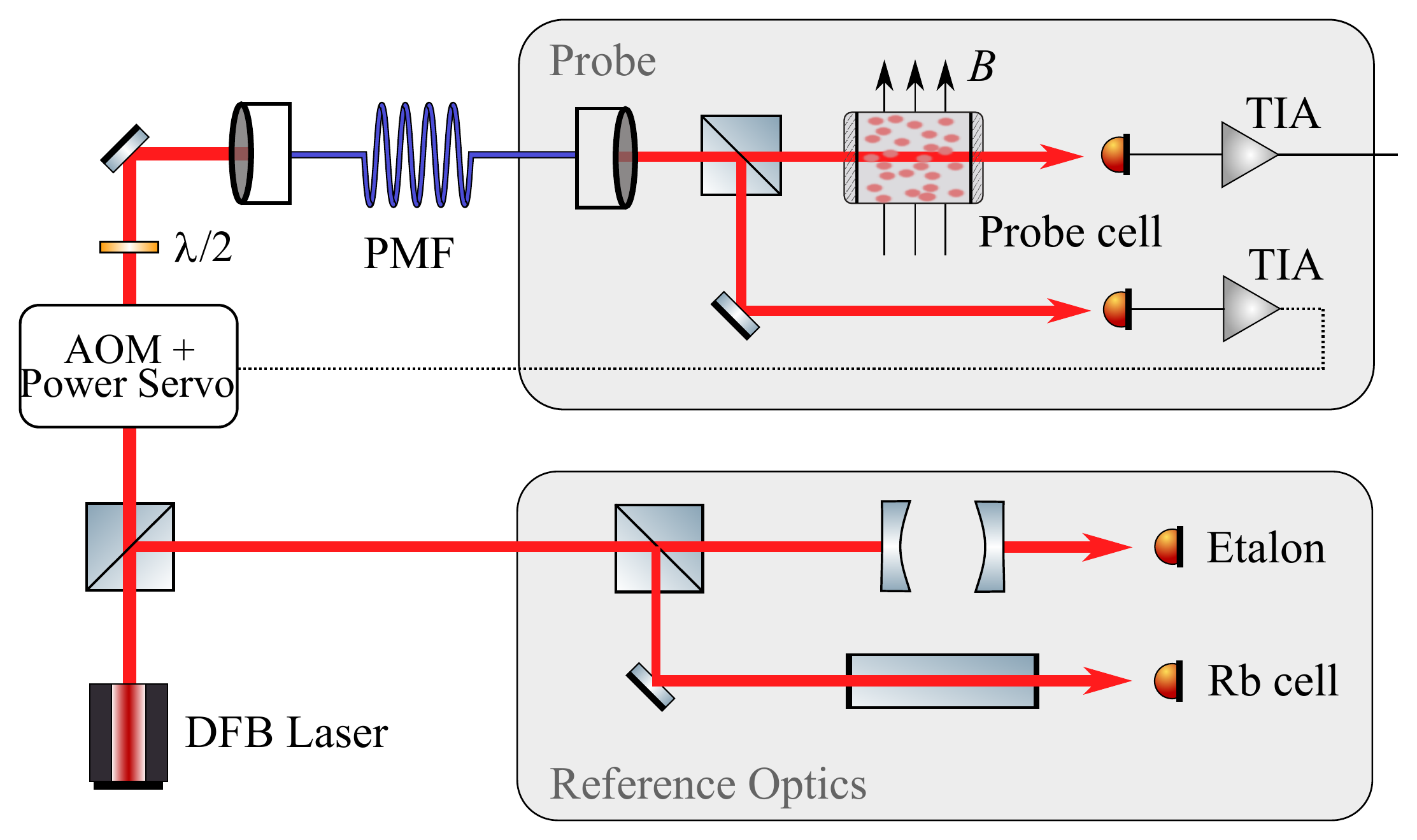}
\includegraphics[width=\columnwidth]{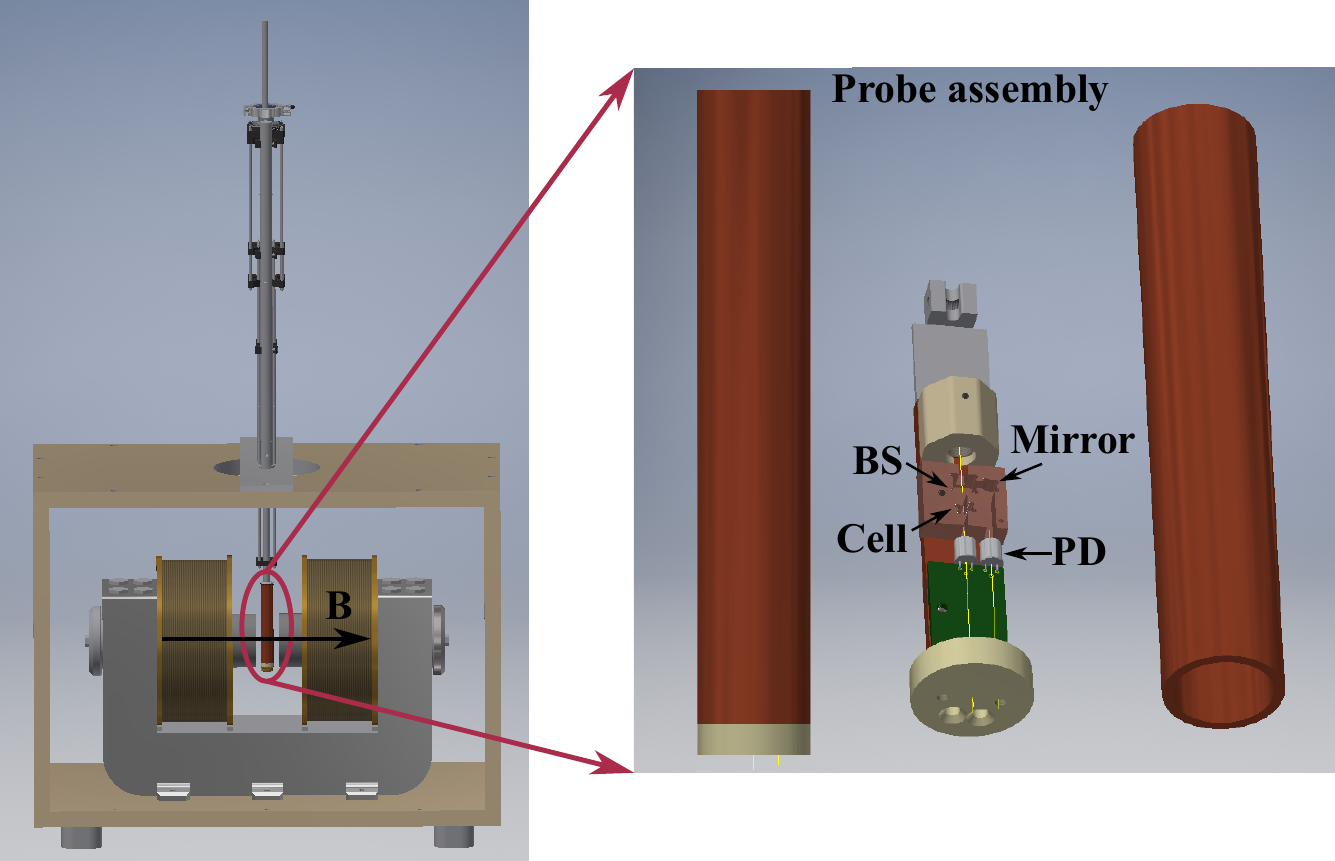}
\caption{Simplified schematic of the optical setup. Light from a distributed feedback (DFB) laser is split into two parts. One part is sent to reference optics, which is a combination of a Fabry-Perot etalon and a 75~mm natural abundance Rb reference cell and is used to calibrate the laser scan in zero magnetic field. The remaining light is sent to the probe via a polarisation maintaining fiber (PMF). The probe itself contains a fiber collimator, after which the light is split by a non-polarising beamsplitter cube. One arm contains the 1~mm $^{87}$Rb isotopically enriched ($99\%\,^{87}$Rb) vapour cell, while the other arm is used as a power reference, which is fed back to a power servo using an acousto-optic modulator (AOM) to keep the power constant~\cite{Truong2012}. Transimpedance amplifiers (TIA) convert the current generated by the photodidodes (PD) into a voltage signal. Below, a detailed render of the experimental setup can be seen showing the position of the probe in the electromagnet, as well as the interior of the probe with the position of the beamsplitter cube (BS), the vapour cell and photodiodes indicated. Measurements of the magnetic field profile were taken by removing the probe and employing a commercial Hall probe.}
\label{fig:setup}
\end{figure}
%

\begin{figure*}[tb] 
\includegraphics[width=2.0\columnwidth]{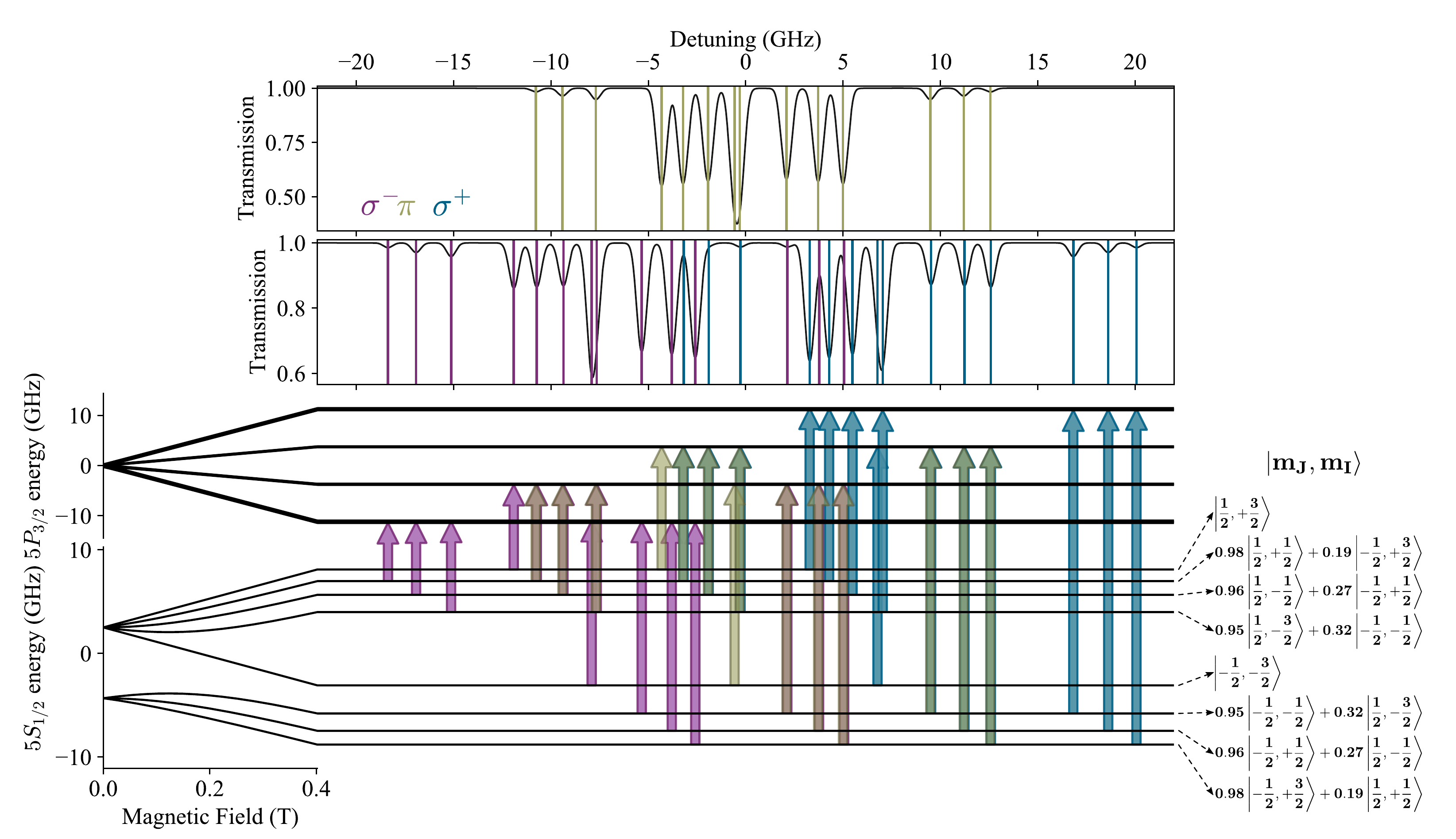}
\caption{Spectroscopy of the alkali D-lines in the Voigt geometry for a magnetic field strength of 0.4~T. The left side of the diagram (at the bottom) shows the evolution of the 5S$_{1/2}$ and 5P$_{3/2}$ atomic energy levels of $^{87}$Rb with magnetic field, up to 0.4~T. The right side shows the states involved in the atomic transitions at 0.4~T. The upper panel shows the calculated spectrum for $\pi$ (olive) transitions ($\vec{E} \parallel \vec{B}_{\rm{ext}}$), while the lower panel shows a calculated spectrum for the $\sigma^+$ (blue) and $\sigma^-$ (purple) transitions ($\vec{E} \perp \vec{B}_{\rm{ext}}$). The coloured arrows indicate the initial and final states involved.}
\label{fig:bigdiagram}
\end{figure*}

\section{Experimental setup}
\label{sec:expt}

The experimental setup is shown in figure~\ref{fig:setup}.
Using either the D1 or D2 line was possible, but the D1 line has a weaker transition strength despite having a simpler hyperfine structure.
As such, the laser chosen for this experiment only allows us to access the D2 line.
The laser is a 780~nm distributed feedback (DFB) laser with a quoted linewidth of $<2$~MHz, which is tunable without mode hops over many hundreds of GHz.
Some of the light is sent to reference optics to calibrate the laser scan; we use a Fabry-Perot etalon to linearise the scan in parallel with a 75~mm natural abundance Rb reference cell which provides an absolute frequency reference, using the approach outlined in ref.~\cite{Keaveney2014a}.
The remaining light is sent to the probe through an acousto-optic modulator (AOM) which we use in conjunction with a power servo to stabilise the laser power following the method in ref.~\cite{Truong2012} and then along a polarisation-maintaining single-mode optical fiber.

In the probe the light is re-collimated, and further split into two; one arm is a reference channel which acts as the feedback signal for the AOM and power servo, and the other arm is the signal channel. 
The signal channel consists of a 1~mm long isotopically enriched ($99\%\, ^{87}$Rb) vapour cell, as used in ref.~\cite{Zentile2014a}, which is heated to provide the required atomic density and hence optical depth.
Fabrication details of the vapour cell can be found in ref.~\cite{Knappe2005} and details of the isotopic purity and spectral qualities can be found in~\cite{WellerThesis2013}.
Both arms are mounted on a central bed of copper wich also includes an internal heater, and a commercially calibrated magnetic-field-insensitive Cernox resistance thermometer (CX-1070-SD-HT-4M)~\cite{Brandt1999} that is $<5$~mm from the vapour cell. 
The copper bed is surrounded by a copper shield that is only in weak thermal contact with it.
The copper shield has the external heaters mounted onto it and is surrounded by a layer of Aerogel insulation.
A detailed view of the probe can be seen in the bottom part of figure 2 showing the previously mentioned elements and the setup within the electromagnet.

In operation the background temperature is set by holding the power in the external heaters to be constant.
Temperature stability is then maintained by using the internal heaters and Cernox thermometer in feedback via a temperature controller.
This maintains a stability of better than 100~mK over the course of an experimental data run.
In these experiments the external magnetic field is generated by an electromagnet with the field in the horizontal direction, and measured using a commercial Hall probe (Hirst GM04).
The field is not perfectly linear with the current, but gives a remanent field of $(10.0 \pm 0.5)$~mT.
Nevertheless, field strengths up to 0.4~T are reproducible to better than 1~mT, with a uniformity better than 1~mT over a 14~mm diameter-sphere-volume.

To avoid optical pumping, the optical power in the probe is kept low such that the atoms are in the weak-probe regime~\cite{Sherlock2009}; in practice this means around 1~$\mu$W of optical power with a beam waist of around 0.5~mm.
The effective spatial resolution of the field probe is then set by the volume of atoms interrogated by the laser beam, which roughly comprises a cylinder of length 1~mm and radius 0.5~mm.

\section{Spectroscopic analysis for magnetic field determination}
\label{sec:spectra}

Figure~\ref{fig:bigdiagram} is derived using our theoretical model \textit{ElecSus} and combines the atomic state evolution as a function of magnetic field with the spectroscopic signals at an external magnetic field of 0.4~T. 
The top two panels show the calculated spectrum for an isotopically enriched ($99\%$ purity) vapour cell at $80^{\circ}$C and a field of 0.4~T for $\pi$ (upper panel) and $\sigma^\pm$ (lower panel) transitions.
The 1$\%$ $^{85}$Rb impurity content in our isotopically-enriched cell is taken into account in the calculations but is sufficiently small that it has no noticeable effect on the spectra.
Coloured vertical lines indicate the frequency that the resonance lines occur at, while the colour of the line indicates the type of transition.
Zero probe detuning is the weighted D2 line centre of naturally abundant rubidium in a zero magnetic field \cite{Siddons2008b}.
Below these panels, the diagram shows the states in the 5S$_{1/2}$ and 5P$_{3/2}$ manifolds at 0.4~T and the initial and final states involved in each individual transition.
The state decompositions in the $m_{J},m_{I}$ basis are shown on the right side of the figure.
The arrows here are semi-transparent to highlight where there are still partially overlapping transitions. 

\begin{figure}[tb] 
\includegraphics[width=1.0\columnwidth]{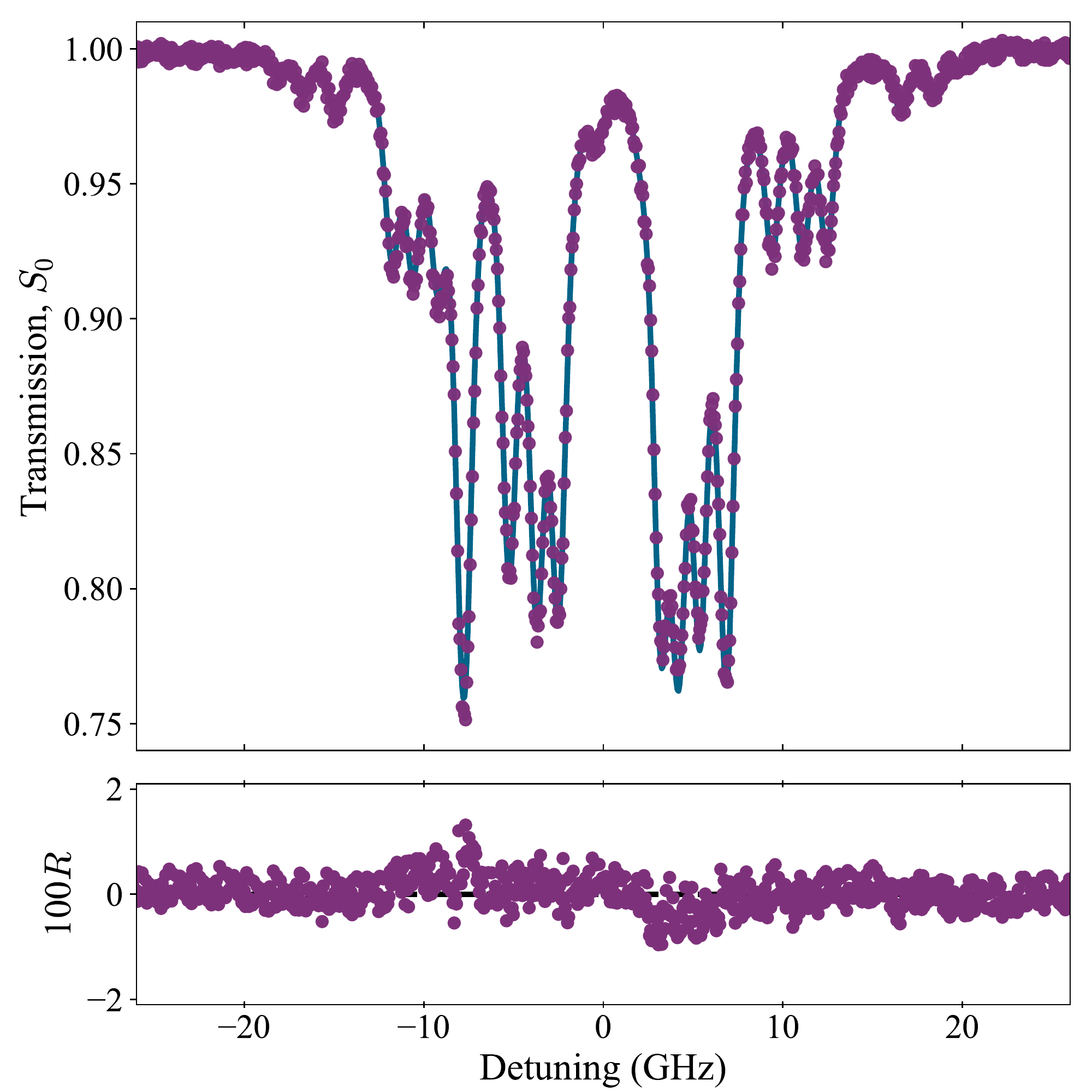}
\caption{Experimental data and fit to the {\it ElecSus} model for a measured magnetic field strength $(390\pm1)$~mT, and average polarisation angle $\phi_{B}=(\pi/2\pm0.02)$. The residuals $R$ (difference between experiment and theory, multiplied by a factor of 100 for clarity) are very small and have no clear structure, which combined with the small RMS error of 0.3\% indicate excellent agreement between theory and experiment. From this fit, we extract a magnetic field strength $|B_{\rm{atoms}}| = (394 \pm 4)$~mT, temperature $T_{\rm{atoms}} = (81.23\pm0.02)^{\circ}$C, polarisation angle $\phi_{B} = (1.412\pm0.004)$ rad and inhomogenous broadening $\Gamma_{\rm{Buf}}/2\pi = (631\pm3)$~MHz.}
\label{fig:elecsus_fit}
\end{figure}

In an applied magnetic field of 0.4~T, the 5P$_{3/2}$ states strongly decouple into the $m_J, m_I$ basis, leading to four groups of lines organised by the $m_J=3/2,1/2,-1/2,-3/2$ projection. 
However, the ground state has a significantly larger hyperfine interaction which means this decoupling is still incomplete.
While all the states have roughly separated into the two $m_J = \pm1/2$ groups, within each group the energy difference between each $m_I$ component is not uniform until the external fields are much higher and the states have evolved completely into the Hyperfine Paschen-Back (HPB) regime. 
Hence, in addition to the groups of four `strong' transitions ($\vert m_J, m_I\rangle \rightarrow \vert m_{J}', m_I\rangle$, with $m_{J}' = m_J \; (\pi), m_J+1 \; (\sigma^+), m_J-1 \; (\sigma^-)$), we also observe groups of three `weak' transitions, which result from the ground-states not being pure eigenstates in the $m_J,m_I$ basis; a small admixture of the opposite $m_J$ state remains which can be seen in the state decomposition on the bottom right of fig.~\ref{fig:bigdiagram} (more details can be found in ref.~\cite{Zentile2014a}).
\begin{figure}[tb]
\includegraphics[width=1.0\columnwidth,clip=true,trim = 0mm 5mm 0mm 2mm]{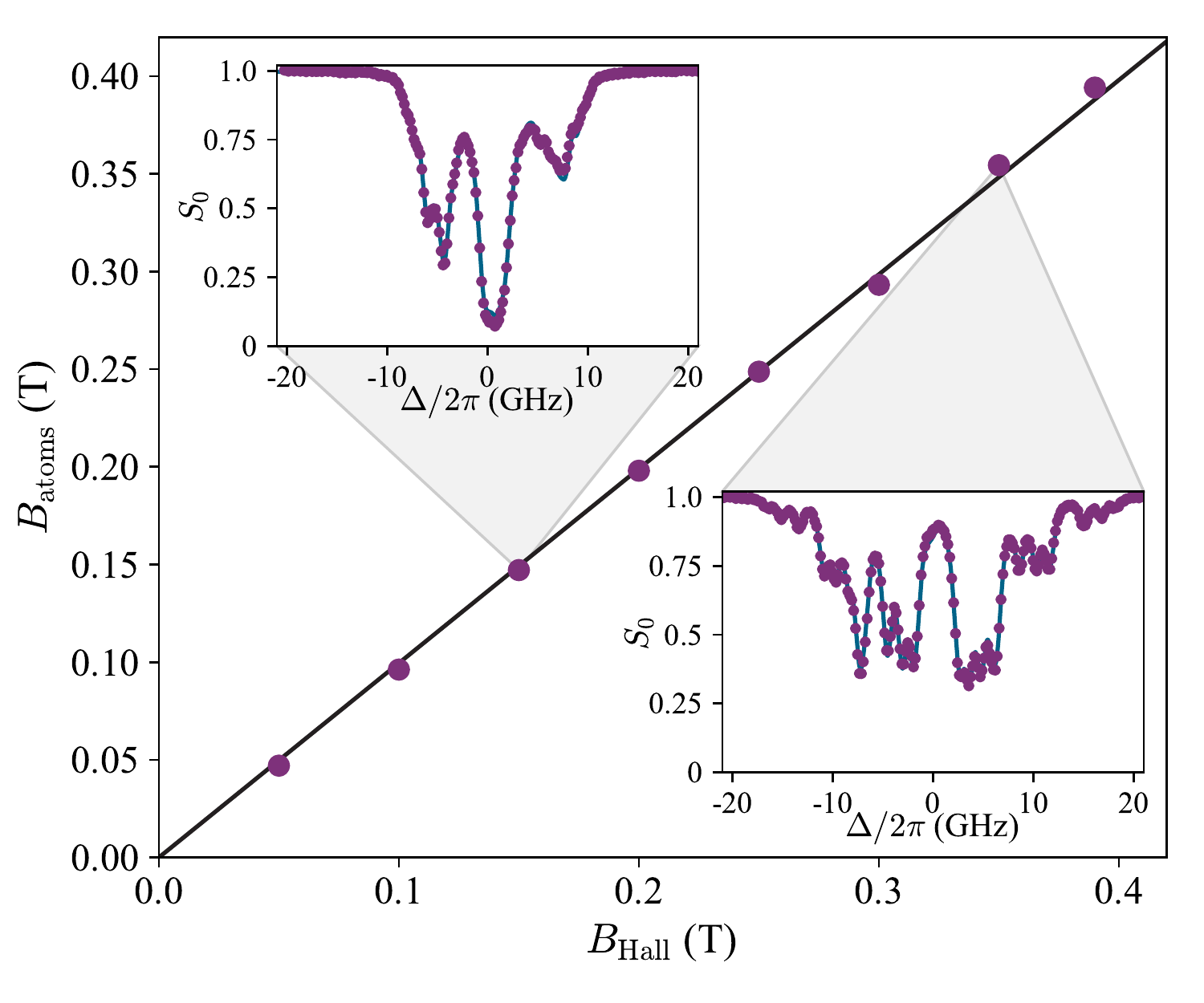}
\caption{Magnetic field strength comparison between measurement using a Hall probe and measurement from fitting spectroscopic data. For each value of magnetic field, we take 5 spectra and plot the average fitted magnetic field. The weighted error bar of each point is approximately 1~mT and is too small to be seen on the figure. Fitting a linear function to the data yields the following: With the intercept fixed to zero, we extract a gradient of $m=(0.995\pm0.009)$. With the intercept floating, we extract $m=(1.015\pm0.002)$ and $c=(-5\pm1)$~mT. The black line in the figure shows the fit with intercept fixed at zero.}
\label{fig:atoms_vs_hall}
\end{figure}
Given that at 0.4~T, the Zeeman splitting is still not large enough to completely resolve all the atomic transitions individually there are areas of the spectrum where lines overlap.
This is most prominent around approximately $\pm7$~GHz in the $\sigma^\pm$ transitions and around -1~GHz for the $\pi$ transitions, where the outer lines from two groups of strong transitions are separated by less than the Doppler width ($\Gamma_{\rm{Doppler}}/ 2\pi = 555$~MHz at $T_{\rm{atoms}}=80^{\circ}$C), and hence the transitions appear merged into one.
While this is not ideal for identifying individual transitions, this may be advantageous for magnetic field detection, as these areas of the absorption spectrum change rapidly with relatively small changes in magnetic field.
We will discuss this in more detail in section~\ref{sec:sensitivity}.
\begin{figure*}[tb] 
\includegraphics[width=1.8\columnwidth]{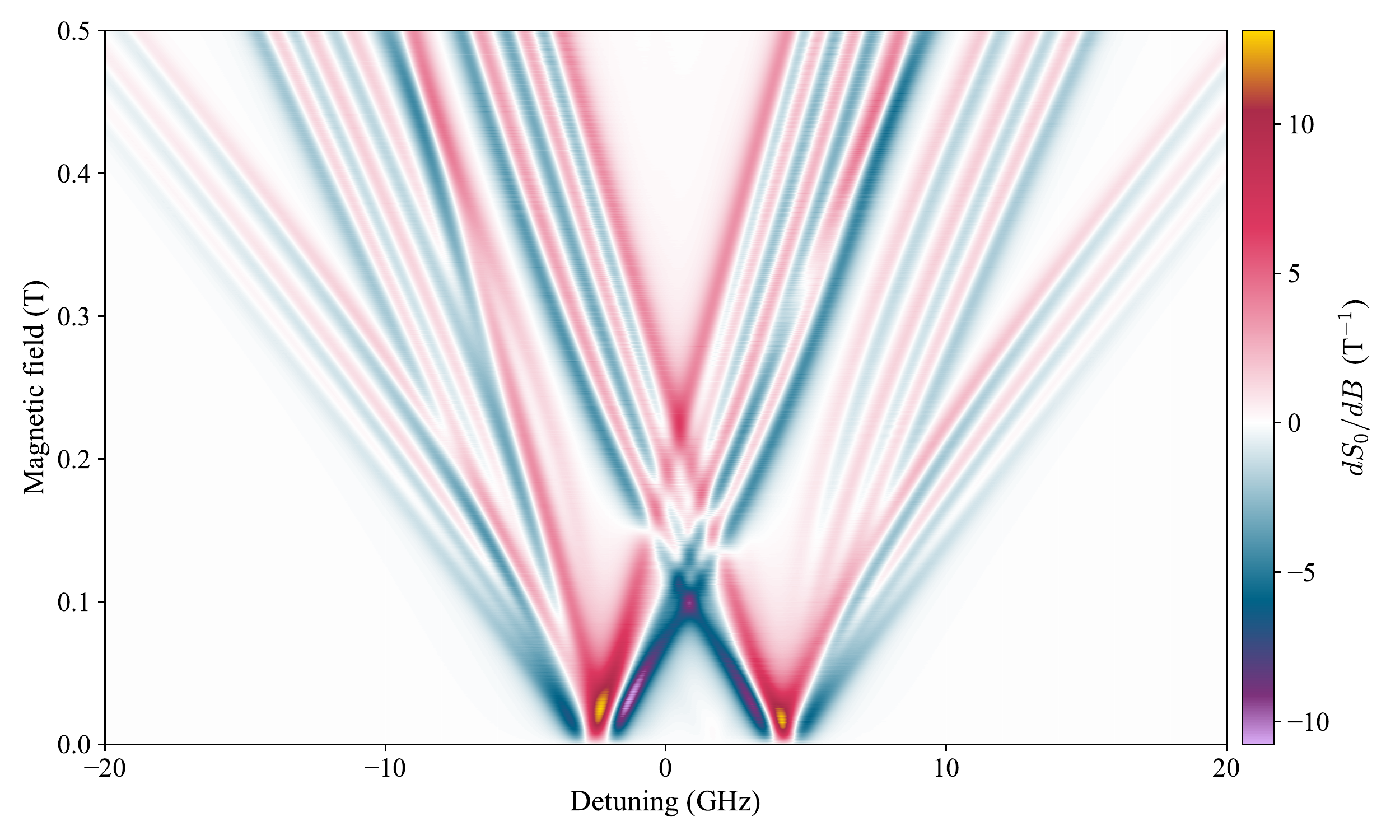}
\caption{Calculated sensitivity of the spectral response with changing magnetic field. We plot the gradient ${\rm d}S_0/{\rm d}B$ as a function of detuning and magnetic field. Hot-spots, indicated by bright purple/yellow areas on the colour map, can be seen where resonance lines overlap, which represent particularly sensitive parts of the spectrum. The calculation uses polarisation angle $\phi_{B} = \pi/2$, cell length $L=1$~mm, $T=80^\circ$C and $\Gamma_{\rm{Buf}}/2\pi=650$~MHz in order to match the experimental parameters as closely as possible.
}
\label{fig:Bsens}
\end{figure*}

Figure~\ref{fig:elecsus_fit} shows experimental data obtained at 390~mT fitted using \textit{ElecSus} along with the residuals $R$, which have been multiplied by a factor of 100 for clarity.
The RMS difference between theory and experiment for this data set is 0.3\%, and combined with the lack of any discernible structure in the residuals, indicates an excellent fit \cite{Hughes2010}.
The fitted parameters are the magnetic field strength $\vert B_{\rm{atoms}} \vert$, the temperature $T_{\rm{atoms}}$, the polarisation angle $\phi_{B}$ and the amount of inhomogeneous broadening $\Gamma_{\rm Buf}$ which is caused by collisions with buffer gas atoms.
We find $\vert B_{\rm{atoms}} \vert =  (394 \pm 4)$~mT, close to the field value measured with a Hall probe of ($390\pm1$)~mT, $T_{\rm{atoms}} = (81.23\pm0.02)^\circ$C which is close to the measured value of $(82.32\pm0.04)^\circ$C from the Cernox thermometer, $\phi_{B}= (1.412\pm0.004)$ rad and $\Gamma_{\rm Buf}/2\pi =  (631\pm3)$~MHz.
These were the only free physical parameters used in our fit, with all remaining elements included in the model generated by \textit{ElecSus}.
The exact origin of the extra buffer gas broadening is unknown.
We suspect the significant buffer gas broadening to be due to He atoms diffusing into the cell over the extended period that it was exposed to a helium-rich gas environment while at elevated temperatures during previous measurements and note there is no significant shift in any of the resonance lines. 

Figure~\ref{fig:atoms_vs_hall} shows detailed measurements and analysis over a range of fields up to 0.4~T.
Each data point is based on the weighted average of 5 fits, examples of which are shown in the insets for an external field of $\sim0.15$~T and $\sim0.35$~T.
We suggest that there will be benefits for very high precision spectroscopy in the HPB regime at even higher fields, where preliminary results show the large Zeeman splitting allows the isolation of individual two-, three- and four-level systems~\cite{Zentile2014a,Whiting2015,Whiting2016a,Whiting2017,Whiting2017b}.

Error bars on the fitted magnetic field strengths are on the order of 1~mT.
The gradient of the $|B_{\rm{Hall}}|$ vs. $|B_{\rm{atoms}}|$ data is linear.
If we fix the intercept to zero, we find a gradient of $(0.995\pm0.009)$. 
Allowing both gradient and intercept to vary, we extract a gradient of $(1.015\pm0.002)$ and intercept $(-5\pm1)$~mT.
These results indicate a systematic 1.5\% difference between the measurements and theoretical fits which we attribute predominantly to inaccuracies in the calibration of the Hall probe, which is quoted to be $\pm1\%$, 1~mT resolution.
Systematic errors in the scaling of the frequency axis, misalignment in the axis of the Hall probe with the magnetic field axis and uncertainties in the theoretical calculations are also considerations.
No magnetic shields were used in the experiment because the effects of the Earth’s field or other sources of parasitic magnetic fields are negligible in our system, typically of $<10^{-4}$, when compared to the applied fields in this work.
We also found no significant changes in temperature or polarisation while these variable-field data were obtained.
The polarisation drift across the measurements was $<10 \%$ and the temperature drift was $<1^{\circ}$~C.
The stability of the cell temperature is consistent with the small changes in the temperature measured by the Cernox of $<100$~mK.

Our atomic technique for measuring fields is independently sensitive to the field strength and angle without any mechanical movement of the sensor head.
The Zeeman shift, which sets the resonance line positions, depends only on field strength, whilst the coupling of the atomic transitions to the light field depends on the relative direction between the magnetic field and the electric field vector of the light.
This is in contrast to Hall probe measurements where the angle between the Hall probe and the direction of the external field must be known to get a precise reading of the magnitude.
Combined with the relatively high spatial resolution, these considerations mean that atomic-based spectroscopy could replace Hall probes for say vector magnetometry or simultaneous mapping of the magnitude and direction of magnetic fields.

%
\section{Spectral sensitivity to changes in field strength and direction}
\label{sec:sensitivity}

In the moderate fields studied here, although the transition peaks are partially overlapping, we argue it may be advantageous for magnetic field sensing, as small changes in the magnetic field can lead to relatively large changes in the spectral lineshape.
In figure~\ref{fig:Bsens} we present calculations using \textit{ElecSus} that show the change in transmission with respect to magnetic field strength, ${\rm d}S_0/{\rm d}B$, over the Rb D2 line spectrum, as a function of field strength.
There are interesting regions of high sensitivity visible around $\sim1$~GHz detuning at 0.1 and 0.23 T.
\begin{figure*}[t] 
\includegraphics[width=2\columnwidth]{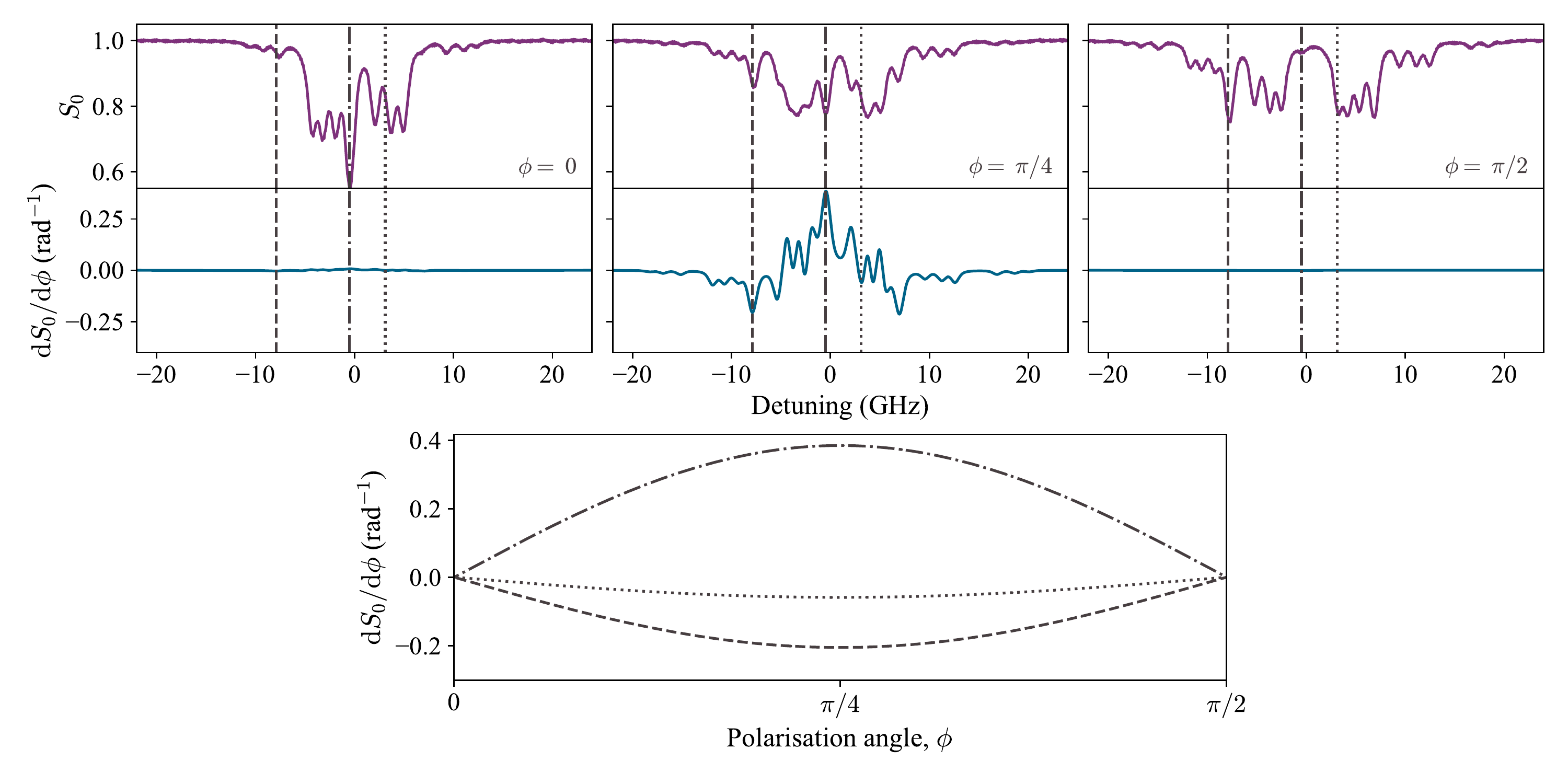}
\caption{Spectral changes with polarisation, and relative sensitivity. We show the three extreme cases of $\phi_{B} = 0, \pi/4, \pi/2$. The top panels show experimental data (purple) at each value of $\phi_{B}$, with the same magnetic field strength of $390$~mT, and a cell temperature $T=80^\circ$C. Below the experimental data, we show the calculated gradient ${\rm d}S_0/{\rm d}\phi_{B}$ in blue. The three panels have the same vertical axis, to show the contrast between the extremely insensitive values $\phi_{B} = 0,\pi/2$ and the extremely sensitive $\phi _{B}= \pi/4$. Dashed vertical lines mark the detunings used to calculate the sensitivity change with angle shown on the bottom panel.
}
\label{fig:Phi_sens}
\end{figure*}

The best sensitivity to magnetic field strength uses a polarisation angle $\phi_{B}=\pi/2$, because this is the angle which drives the $\sigma^\pm$ transitions the strongest.
These transitions have a larger energy splitting  as can be seen in figure~\ref{fig:bigdiagram} which shows that in the HPB regime the ground state level shifts down in energy by $\mu_{\rm{B}}B_{\rm{ext}}$ while the excited state level shifts up by $2\mu_{\rm{B}}B_{\rm{ext}}$.
The spectra are also dependent on the relative angle between the magnetic field vector $\vec{B}_{\rm{ext}}$ and the light's electric field vector $\vec{E}$.
Whilst this technique gives two equivalent polarisation angles (i.e. reflection of the polarisation angle around the $x$ and $y$-axes yield the same spectrum), a full polarimetric analysis would in principle be able to determine the field angle uniquely.
In figure~\ref{fig:Phi_sens} we plot experimental spectra (top panels, purple) at three polarisation angles, and the change in transmission with respect to polarisation angle ${\rm d}S_0/{\rm d}\phi_{B}$ over the spectrum for each of these cases.
Since the polarisation angle changes the relative coupling to the $\pi$ and $\sigma^\pm$ transitions according to $\cos^2(\phi_{B})$ and $\sin^2(\phi_{B})$, respectively, the spectrum is most sensitive at an angle of $\pi/4$, where the gradient of both functions are largest.

\begin{figure}[H]
\includegraphics[width=\columnwidth]{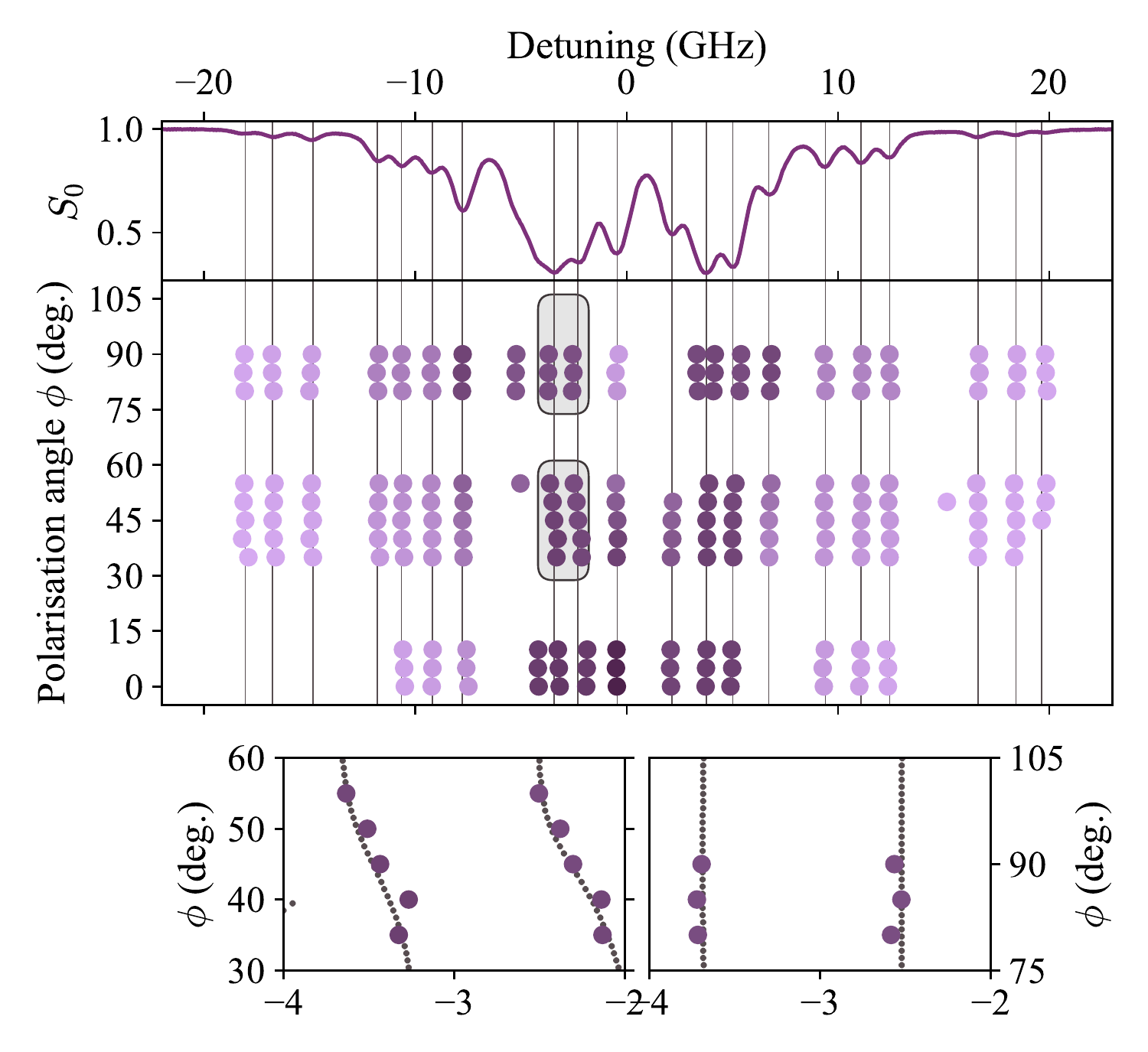}
\caption{Position of spectral local minima with changing $\phi$. The top panel shows an experimental spectrum at $\phi=\pi/4$ for reference. The main panel shows the position of local minima in the spectra for a range of polarisation angles. The colour of the points represents the transmission (dark colours are low transmission). The insets show a zoom in of the two highlighted regions, which demonstrates the relative sensitivity of the local minima positions as the polarisation angle is changed.}
\label{fig:domino}
\end{figure}

An alternate demonstration of the polarisation sensitivity is shown in figure~\ref{fig:domino}.
Here we plot positions of the local minina of a series of experimental spectra at different polarisation angles, whilst maintaining the magnetic field value at a constant value of 395~mT.
The colour of the points denotes the value of transmission at the local minina - dark shades are low transmission, light shades are high transmission.
The top panel is for reference and shows a spectrum at $\phi_{B}=\pi/4$, and the solid vertical grey lines are the resonance positions in this spectra, for comparison.
The two bottom panels show the shaded regions in the main plot, and highlight the relative sensitivity and insensitivity of the local minina.
The grey dashed line shows the theoretical positions of the local minima.
The scatter of the experimental data about this curve reflects a combination of uncertainty in the local minima position due to the finite signal-to-noise ratio and small fluctuations in the magnetic field value between data runs.
Because of the overlap of spectral lines, changing the relative coupling strength of these transitions also changes the positions of the local minima, and is most prominent when $\phi_{B}$ is close to $\pi/4$. When $\phi_{B}$ is close to 0 or $\pi/2$, the positions are much less sensitive to changes in $\phi_{B}$, as evidenced by the near-vertical lines.

It should be noted that the relative sensitivity to either magnetic field strength or polarisation angle changes with the line width of the spectral features.
The narrower the features, the sharper the change and hence a more sensitive measurement can be made.
One recent development in spectroscopic techniques is that of so-called `Derivative of Selective Reflection' (DSR) in ultra-thin vapour cells~\cite{Klinger2017,Sargsyan2017a}, which is a modulation-free method to increase spectral resolution over conventional transmission and selective-reflection spectroscopy.
Whilst modelling the spectra is more complex than in transmission spectroscopy, the use of sub-micron thickness vapour cells also increases the effective spatial resolution, making these systems very promising for future research.

\section{Conclusions}
\label{sec:conclusions}
In conclusion, we have demonstrated a technique to measure the absolute magnetic field strength and angle of polarisation using a thermal vapour of alkali-metal atoms.
Our results use $^{87}$Rb, but the technique is applicable to any alkali-metal atom.
We have found excellent agreement between our detailed spectroscopic data and our theoretical model of the transmission through the medium.
Some preliminary results also demonstrate that this experimental approach may be successful up to much higher magnetic fields, which will be the subject of future work.
One can also envisage incorporating polarimetric techniques to fully constrain the polarisation angle, but this will bring with it several new challenges, some foreseen, e.g., the birefringence in the vapour cell windows~\cite{Keaveney2017a}, and others not.

\section{Acknowledgements}
The authors would like to thank Stephen Lishman and members of the workshop for their aid in fabricating the mechanical components for the experiment, and acknowledge funding from EPSRC (Grant Nos. EP/L023024/1 and EP/R002061/1) and Durham University. FSPO gratefully acknowledges a Durham Doctoral Scholarship. The data presented in this paper is available from \footnote{\url{doi.org/10.15128/r2sq87bt617}}.
\vfill
\pagebreak

%

\bibliography{High_field_magnetometry_resubmission2}

\end{document}